\documentclass[runningheads]{llncs}

\usepackage[T1]{fontenc}

\usepackage{graphicx}
\usepackage{tikz}
\usetikzlibrary{positioning,fit,arrows.meta}

\usepackage{xcolor}

\begin{document}
\title{AI in the Wild: A Large Scale Analysis of Authentic Interactions of College Students with Generative AI}

\author{
Taelin Karidi\inst{1}\orcidID{0009-0008-3368-7394} \and
Ofra Amir\inst{1}\orcidID{0000-0003-2303-3684} \and
Ido Roll\inst{1}\orcidID{0000-0001-7295-9059}
}
\authorrunning{Karidi et al.}
\titlerunning{AI in the Wild: Student--AI Interaction Patterns}

\institute{
Technion - Israel Institute of Technology, Haifa, Israel \\
\email{taelinkaridi@campus.technion.ac.il, oamir@technion.ac.il, roll@technion.ac.il}
}

\maketitle              
\begin{abstract}

Generative AI tools (GenAI) are increasingly used by students during coursework, yet empirical understanding of how students engage with these systems in authentic learning contexts remains limited. Existing studies have largely relied on controlled settings, single-domain analyses, or small-scale qualitative data, leaving open how student-AI interaction unfolds across courses and forms of academic work.

We present a large-scale analysis of naturally occurring student-AI interactions collected from undergraduate students across multiple university courses and academic domains. The dataset comprises over 15{,}000 student-AI interaction units drawn from voluntary use of generative AI during real coursework.

To characterize these interactions, we analyze each student turn along two complementary dimensions, cognitive intent and interaction context, capturing whether requests are directed toward the task or domain, the student’s own work, or prior AI output. Using instruction-guided annotation applied at scale, we examine how these interaction patterns are distributed overall and how they vary across courses.

Our analysis reveals that student-AI interaction is highly structured. Across courses, interactions concentrate in a small number of recurring patterns rather than exhibiting highly idiosyncratic use. At the same time, systematic differences emerge across courses, giving rise to distinct interaction profiles associated with different forms of academic work. 

\end{abstract}

\keywords{AI  \and Learning Analytics \and Human-AI Interaction.}

\section{Introduction}
\label{sec:intro}

GenAI tools, and large language models in particular, are increasingly embedded in students’ everyday academic use \cite{wang2025effect,bilal2025guest}. Students use these systems across learning activities, from problem solving to writing, and across academic domains \cite{haindl2024students,levine2025students}.

Despite this growing presence, empirical understanding of how students engage with GenAI during authentic learning activities remains limited. 
Existing work has largely focused on controlled settings, single-domain scenarios, or small-scale qualitative studies, offering detailed insight into particular behaviors but leaving open how student–AI interaction unfolds across courses and forms of academic work \cite{han2023exploring}.
While such studies provide valuable insight into specific behaviors and outcomes, they constrain how students can interact with AI systems, and thus offer only a partial view of how generative AI is incorporated into students’ learning practices when the use is voluntary and shaped by real task demands.

In this paper, we study student–AI interaction in the wild, using a large corpus of naturally occurring interactions collected across multiple university courses and academic domains. We conceptualize each student turn along two complementary dimensions: the cognitive intent underlying the request (Bloom level), and the interaction context to which it is directed. The former captures the learning objective expressed in the student’s message, while the latter characterizes how students position the AI relative to their own work, the task itself, or prior AI output. Using an instruction-guided annotation framework applied at scale, we analyze over 15,000 student–AI interaction units, collected from authentic coursework, examining variation across courses, within conversations, and in students’ engagement with AI-generated responses.

Our analysis reveals that despite the breadth of tasks and domains, student–AI interaction concentrates in a small number of recurring patterns.
These patterns are shared across many students and conversations, yet differ systematically in how they are realized across courses. These systematic differences between courses give rise to distinct interaction profiles that reflect differences in academic work rather than idiosyncratic individual behavior.

\section{Background}
\label{sec:background}

Educational research has repeatedly shown that the learning value of an interactive tool depends on \emph{how} learners use it, not only on whether students have access to it \cite{aleven2003,roll2023measuring}. Students' use of available resources shows much variability and is linked to differences in learning outcomes \cite{drive2025towards,VanLehn2011}. More broadly, learning sciences emphasize that observable learner actions during interaction (e.g., asking for explanations) provide a window into learning processes \cite{chi2023applying}.
Within AI in Education, a long tradition studies interaction patterns in one-to-one tutoring and dialogue settings, including how students ask questions, respond to feedback, and iterate on solutions \cite{graesser2016conversations,VanLehn2011}. These works motivate analyzing student-AI conversations through interaction structure, not only outcomes.

In the context of Large Language Models (LLMs), there is a fast growing interest in how students use tools such as ChatGPT for authentic academic work, but much of the evidence base is still dominated by surveys, small cohorts, or studies in a single domain \cite{vcrvcek2023writing,hao2025ai}. Even log-based studies often remain limited in the number of participants \cite{ammari2025students}. For instance, Ammari et al.\ analyze exported ChatGPT histories from 36 undergraduates. In parallel, domain-specific datasets have started to appear, RECIPE4U reports a semester long EFL writing setting with 212 students and releases, alongside intent labels and writing-revision traces \cite{HanEtAl2024RECIPE4U}. At a larger scale but mostly via self-report, recent higher-education surveys document widespread adoption of GenAI among students \cite{HEPIKortext2024} and the relationship between use and understanding is not obvious \cite{klein2026measuring} . This leaves open basic questions about whether stable interaction profiles recur across academic contexts, and how they vary across types of academic work.

\section{Dataset and Collection}
\label{sec:dataset}

\begin{table}[t]
\centering
\small
\begin{tabular}{lrrrrr}
\hline
\textbf{Course} &
\textbf{Students} &
\textbf{Tasks} &
\textbf{Logs} &
\textbf{Turns} &
\textbf{Turns / log} \\
\hline

English &
19 &
4 &
42 &
124 &
6.2 $\pm$ 6.1 \\

Complex Functions &
275 &
4 &
588 &
3754 &
9.9 $\pm$ 8.9 \\

Fourier Analysis &
269 &
4 &
602 &
3839 &
9.3 $\pm$ 7.9 \\

Intelligent Systems &
55 &
3 &
137 &
1467 &
22.2 $\pm$ 26.9 \\

Organizational Behavior &
56 &
4 &
144 &
2597 &
26.8 $\pm$ 24.2 \\

Probability &
147 &
4 &
565 &
4106 &
10.2 $\pm$ 10.6 \\

\hline
\textbf{Total} &
\textbf{821} &
\textbf{23} &
\textbf{2078} &
\textbf{15{,}887} &
-- \\
\hline
\end{tabular}
\caption{Overview of courses and student participation.
Logs correspond to student-submitted chat files.
Turns per log are reported as mean $\pm$ standard deviation, computed over valid logs only.}
\label{tab:course_overview}
\end{table}


We analyze a large-scale dataset of authentic student-AI interactions collected from undergraduate courses at the Technion – Israel Institute of Technology. The dataset includes 821 students, 2,078 submitted chat logs, and 15,887 interaction turns across multiple domains (see Table~\ref{tab:course_overview}). Students voluntarily submitted chat logs generated while working on course assignments, without instructions on how to use AI tools. Participation was incentivized with minor course credit (up to 2\%), with an alternative assignment for non-users. Submissions were independent of course staff, and all data were anonymized; the study was approved by the institutional review board.\footnote{All materials are available at \url{https://osf.io/z9hqb/}.}

\section{Annotation Framework}
\label{sec:annotation}

To characterize how students engage with GenAI tools during learning, we introduce a \textbf{two-dimensional annotation framework} that captures both the cognitive intent of student requests and the interactional context to which those requests are directed. The framework was informed by established educational taxonomies and developed through iterative inspection of the data and recurring patterns in authentic interactions.

The annotations are assigned at the interaction unit level and are based on the inferred intent expressed in the student’s message. The accompanying AI response is used only as contextual information when the student request is underspecified (e.g., follow-up turns such as ``why?'' ).

\begin{table}[t]
\caption{Cognitive intent labels (Dimension 1) with definitions and examples. Labels reflect the \emph{student’s inferred intent}.}
\label{tab:tax_dim1}
\centering
\small
\begin{tabular}{p{2.1cm} p{5.4cm} p{3.3cm}}
\hline
\textbf{Bloom Level} & \textbf{Description} & \textbf{Example} \\
\hline
\textsc{remember} &
Requests for factual information, definitions, or simple recall, without asking for explanation or reasoning. &
``What is the definition of a determinant?'' \\

\textsc{understand} &
Requests aimed at understanding concepts, reasoning, intuition, or intermediate steps (e.g., ``why'' or ``how''). &
``Can you explain why this formula works?'' \\

\textsc{apply} &
Requests for solving a problem or producing a final answer or result, without requesting explanation or justification. &
``Solve this integral.'' \\

\textsc{analyze} &
Requests to analyze, decompose, or interpret an existing artifact, argument, or text. &
``Summarize the main argument and compare it to my solution.'' \\

\textsc{evaluate} &
Requests to assess correctness, validity, or quality, including verification of claims or checking solutions. &
``Is my solution correct?'' \\

\textsc{create} &
Requests to generate, rewrite, or edit content, including text production, paraphrasing, or formatting. &
``Rewrite this paragraph to be more concise.'' \\
\hline
\end{tabular}
\end{table}


\subsection{Dimension 1: Cognitive Intent (Bloom)}
\label{subsec:dim1}

The first dimension captures the \emph{cognitive intent} underlying each student turn, operationalized using Bloom’s revised taxonomy as revised by Anderson and Krathwohl \cite{AndersonKrathwohl2001}. Each student message is labeled as one of six categories: \textsc{remember}, \textsc{understand}, \textsc{apply}, \textsc{analyze}, \textsc{evaluate}, or \textsc{create}. These labels reflect the \textit{objective} expressed in the student’s request rather than the quality or correctness of the resulting work.
Table~\ref{tab:tax_dim1} summarizes the learning objective labels used in our annotation scheme, including operational definitions and representative examples for each category. 


\subsection{Dimension 2: Interaction Context} 
\label{subsec:dim2}

Bloom’s taxonomy characterizes the cognitive demands expressed in a student’s request, but remains agnostic to the object of that request. In student-AI interactions, requests that occupy the same cognitive level may nevertheless reflect qualitatively different modes of engagement. For example, a request to apply a method may be directed toward solving the task itself, evaluating the student’s own attempt, or interrogating a prior response produced by the AI. 

The second dimension captures the \emph{interaction context}, specifying what the student is directing the AI to engage with. Each interaction unit is labeled as:

\begin{itemize}
  \item \textbf{\textsc{own\_work}}: The student’s own work.
  \item \textbf{\textsc{task\_domain}}: The task or domain itself.
  \item \textbf{\textsc{ai\_output}}: A prior response produced by the AI. 
  \item \textbf{\textsc{administrative}} : a purely administrative function.
\end{itemize}
These distinctions capture differences in how responsibility for problem solving is distributed between the student and the AI. Table~\ref{tab:interaction_context} provides formal definitions and illustrative examples for each context label.

\begin{table}[t]
\caption{Interaction context labels (Dimension 2) with definitions and examples. Labels indicate the \emph{object} to which the student directs the AI.}
\label{tab:interaction_context}
\centering
\small
\begin{tabular}{p{2.4cm} p{5.3cm} p{3.1cm}}
\hline
\textbf{Context label} & \textbf{Description} & \textbf{Example} \\
\hline
\textsc{own\_work} 
& The student refers to their own solution, reasoning, or intermediate work. 
& ``Here is my solution — where did I go wrong?'' \\

\textsc{task\_domain} 
& The student refers to the task or domain content itself, independent of any prior AI response. 
& ``Explain Fermat’s theorem.'' \\

\textsc{ai\_output} 
& The student refers explicitly to a previous response produced by the AI. 
& ``In your answer above, why did you assume independence?'' \\

\textsc{admin} 
& The turn contains only administrative or backchanneling content, without substantive task-related intent. 
& ``Thanks'', ``continue''. \\
\hline
\end{tabular}
\end{table}


\section{Annotation Procedure}

Annotations were produced using an instruction-guided LLM following a human-in-the-loop procedure designed to ensure consistent application of the proposed taxonomy at scale \cite{tan2024large,calderon2025alternative}. The annotation prompt was iteratively developed by applying it to subsets of the data, inspecting outputs and refining label definitions and disambiguation rules until stable behavior was achieved. The finalized prompt was then applied uniformly to the full dataset using \textsc{gpt-5-mini}.

Each student turn was annotated based on the inferred intent in the student message, with access to the preceding AI response only to resolve underspecification. The model was constrained to select labels from predefined sets and to output a confidence score for each annotation.

Full annotation guidelines, prompt templates and implementation details are provided in the online repository.

\section{Results}
\label{sec:results}

We perform the analysis at multiple levels of granularity. We first characterize overall distributions of cognitive intent and interaction context across all student turns. Next, we analyze how these interaction patterns vary between courses.

\begin{figure}[t]
  \centering
  \includegraphics[width=0.9\textwidth]{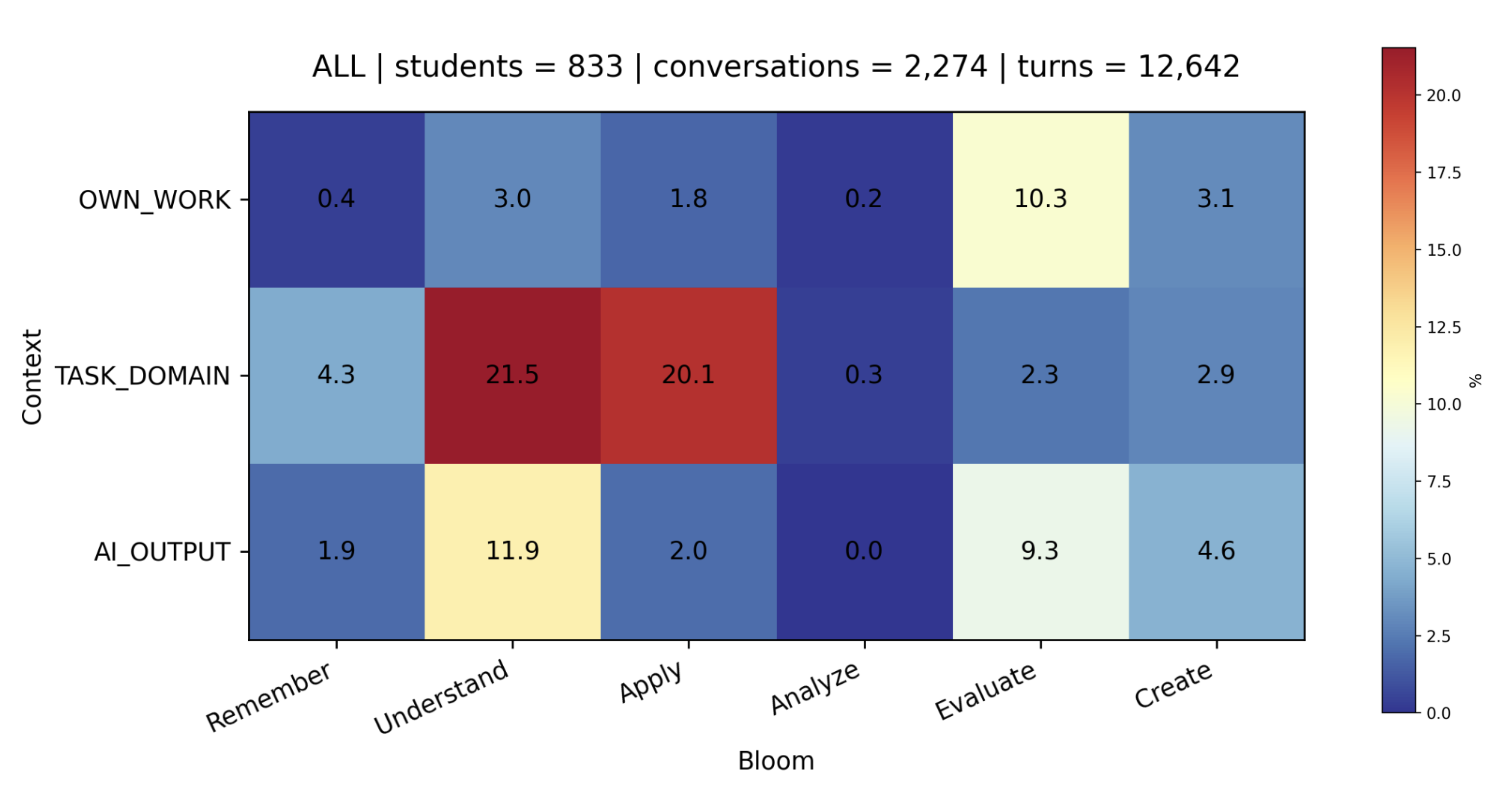}
  \caption{
  \textbf{Overall interaction structure.}
  Joint distribution of interaction context and Bloom-level cognitive intent aggregated across all courses.
  Values represent conversation-normalized percentages, such that each submitted chat log contributes equally, and sum to 100\%.
  }
  \label{fig:context_bloom_overall}
\end{figure}

\subsection{Overall Interaction Structure}
\label{subsec:results_overall}

Across the dataset, interaction mass is concentrated in a small number of recurring patterns (Figure~\ref{fig:context_bloom_overall}). Turns directed toward the task or domain (\textsc{task\_domain}) account for roughly half of interactions, followed by references to AI output (\textsc{ai\_output} $\sim$ 30\%) and to students’ own work (\textsc{own\_work} $\sim$ 20\%). 

In terms of cognitive intent, most interactions occur at the \textsc{understand} and \textsc{apply} levels, with \textsc{evaluate} and \textsc{create} forming a substantial minority, and \textsc{analyze} appearing rarely.

Examining the joint distribution reveals a structured association between context and cognitive intent: task-directed interactions are concentrated at lower to mid Bloom levels, while references to students’ own work are more often associated with higher-level requests. Interactions with prior AI output span multiple levels, but are less common for \textsc{apply} and \textsc{analyze}.

\subsection{Variation Across Courses}
\label{subsec:results_courses}

\begin{figure}[t]
  \centering
  \includegraphics[width=\textwidth]{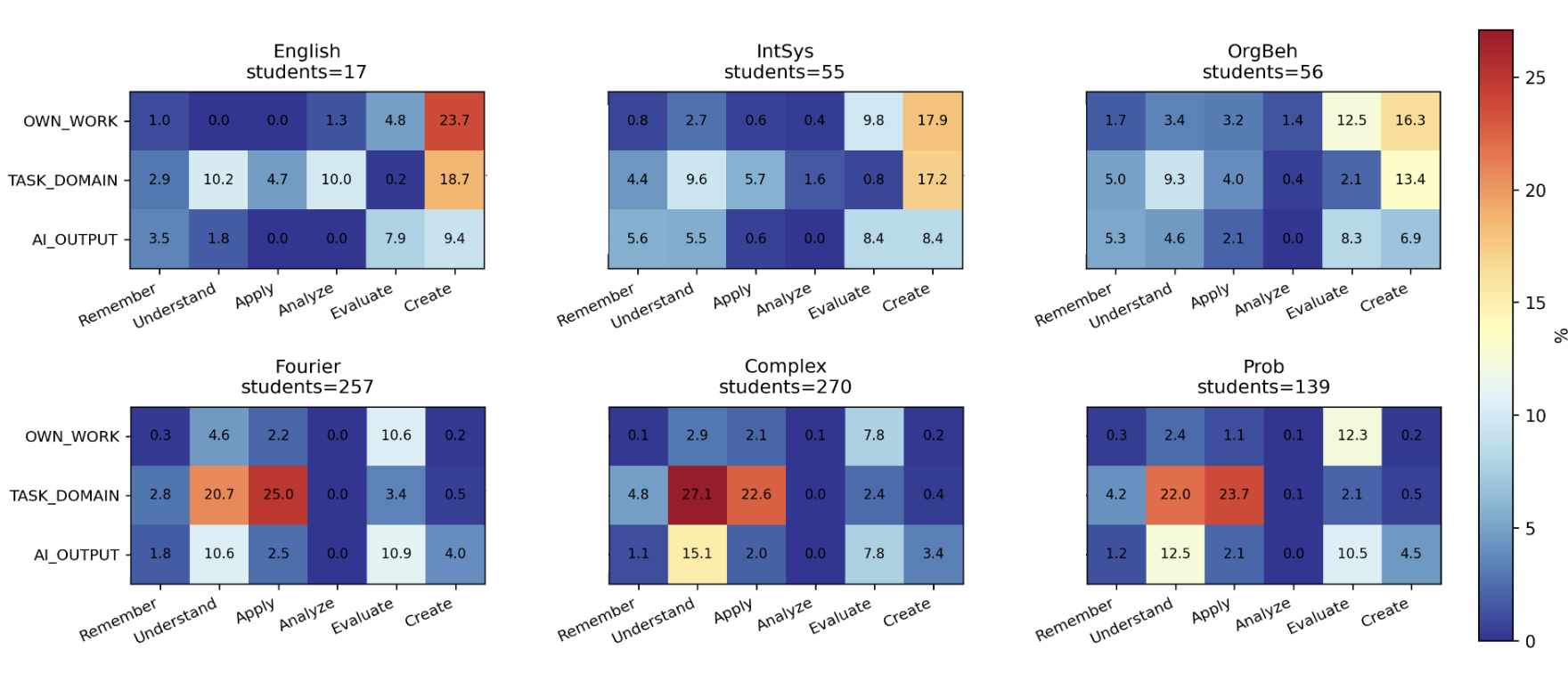}
  \caption{
  \textbf{Interaction Structure Across Courses. }Joint distribution of Bloom-level cognitive intent and interaction context across courses.
  Values are normalized by conversation, such that each conversation contributes equally regardless of length. 
  }
  \label{fig:context_bloom_by_course}

\end{figure}

Figure~\ref{fig:context_bloom_by_course} shows that, despite domain differences, interaction patterns are concentrated in similar regions across courses, suggesting a limited set of recurring interactional forms.

Within this shared structure, course-level differences emerge. Mathematically oriented courses (Fourier, Complex Functions, Probability) are dominated by task-focused interactions at \textsc{understand} and \textsc{apply}, with relatively few references to students’ own work. In contrast, courses involving open-ended writing (English, Organizational Behavior, Intelligent Systems) exhibit more frequent references to students’ own work and a greater presence of higher-level cognitive activity (\textsc{evaluate}, \textsc{create}). These courses also tend to involve longer interactions on average (Table~\ref{tab:course_overview}).

\section{Discussion and Conclusion}

Student–AI interactions are not arbitrary, but concentrate in a small number of recurring patterns across and within courses, suggesting that despite the flexibility of generative AI tools, students tend to use them in a relatively small number of predictable ways. 
This finding extends earlier qualitative and small-scale studies of student use of generative AI \cite{ammari2025students} by showing that similar regularities persist at scale and across academic contexts.

At the same time, the distribution differs systematically across courses, giving rise to systematic differences in how interaction patterns are realized across courses. These differences suggest that interaction patterns are sensitive to the structure of the underlying task. Mathematically oriented courses are dominated by task focused interactions at lower-mid Bloom levels, whereas courses involving open-ended writing exhibit more frequent references to students’ own work and higher-level cognitive activity. This aligns with prior work showing that task structure constrains how learners engage with support tools \cite{KarabenickNewman2006,VanLehn2011}.

We interpret this contrast as reflecting differences in how responsibility for coursework is distributed between students and the AI across domains. In more structured problem solving settings, students tend to use the AI in a supporting role, whereas in open-ended tasks interactions more often engage the AI directly with the content of the work. This interpretation is grounded in the observed distributions and qualitative inspection of the data, linking these interaction patterns to learning outcomes remains an important direction for future work.

Finally, the coding scheme used in this study proved informative in capturing both consistency and variation in student-AI interaction across courses.
Each course exhibits a distinct interaction profile, suggesting that relatively coarse-grained labels can nevertheless reveal meaningful regularities in authentic interaction data.
Future work may link these patterns to course attributes, refine the labeling scheme, or examine how interaction profiles evolve over time.

Taken together, these findings show that student-AI interaction in the wild is neither arbitrary nor purely individual.
Instead, it is structured and systematically shaped by the forms of academic work students are asked to perform. Large-scale, interaction-level analyses of this kind provide a foundation for understanding how generative AI is being integrated into learning practices, and for informing future discussions of instructional design and AI-supported learning.

\begin{credits}
\subsubsection{\ackname}
Funded by the European Union (ERC, Convey, 101078158). Views and opinions expressed are however those of the author(s) only and do not necessarily reflect those of the European Union or the European Research Council Executive Agency. Neither the European Union nor the granting authority can be held responsible for them.
\end{credits}

\section{Limitations and Future Work}
\label{sec:limitation}

Several limitations should be noted. Participation was voluntary, raising possible selection effects: students who submitted chat logs may differ from those who did not. We do not evaluate agreement between LLM-based and human annotations, relying instead on a prompt-based procedure for large-scale analysis; future work could incorporate targeted human annotation. Finally, the study covers a limited number of courses, and extending it would allow assessing generality.

\bibliographystyle{splncs04}
\bibliography{reference}
\end{document}